\documentclass[prl,aps,showpacs,twocolumn]{revtex4} 

\newcommand\be{\begin{eqnarray}}
\newcommand\ee{\end{eqnarray}}
\newcommand\nn{\nonumber}
\usepackage{graphicx} 
\usepackage{epsfig} 
\usepackage{amssymb}
\usepackage{amsmath}
\usepackage{amsfonts}

\begin{document}
\title{Sagnac interferometry based on ultra-slow polaritons in cold atomic 
vapors}
\author{F. Zimmer and M. Fleischhauer}
\affiliation{Fachbereich Physik der Technischen 
 Universit\"{a}t Kaiserslautern, D-67663\\
Kaiserslautern, Germany}
\date{\today }
\pacs{Pacs numbers}
 
\begin{abstract}
The advantages of light and matter-wave Sagnac interferometers --
large area on one hand and high rotational sensitivity per
unit area on the other -- can be combined utilizing ultra-slow 
light in cold atomic gases. While a group-velocity
reduction alone does not affect the Sagnac phase shift, the associated 
momentum transfer from light to atoms generates a coherent matter-wave
component which gives rise to a substantially enhanced rotational 
signal. It is shown that matter-wave sensitivity 
in a large-area interferometer can be achieved if an optically dense vapor 
at sub-recoil temperatures is used. Already a noticeable enhancement of the
Sagnac phase shift is possible however with much less
cooling requirements.
\end{abstract}
\maketitle

The relative phase accumulated by two counter-propagating waves in a rotating
ring interferometer, first observed by Sagnac \cite{Sagnac-1913}, 
is a well established tool to detect intrinsically 
the rotation of a system. 
Two types of Sagnac gyroscopes have been developed which have
very different characteristics depending on the type of wave phenomena 
they are based on \cite{Stedman-RPP-1997}. Optical devices 
achieve high rotational sensitivity
because of the large interferometer area \cite{Chow-RMP-1985}. On the other
hand, the rotational sensitivity per unit area of 
matter-wave gyroscopes \cite{Colella-PRL-1975,Riehle-PRL-1991}
exceed that of optical ones by the ratio $mc^2/\hbar\omega\sim 10^{11}$.
They suffer, however, from the smallness of the achievable loop areas and only
recently short-time rotational sensitivities have been 
demonstrated for matter-wave interferometers which are
comparable to state-of-the-art 
laser gyroscopes \cite{Gustavson-PRL-1997,McGuirk-PRL-2000}. 
We here show that it should be possible to combine the large rotational 
sensitivity per unit area of matter-waves and the large area typical 
for optical gyroscopes utilizing the coherence and momentum transfer 
associated with ultra-slow light in cold atomic vapors
with electromagentically induced transparency (EIT)
\cite{Harris-PhToday-1997,slow-light}.
As the slow-down of light in EIT media is based on the
rotation of dark-state polaritons from electromagnetic to
atomic excitations \cite{Fleischhauer-PRL-2000}, 
light waves can coherently be transformed into
matter-waves, if the excitation transfer is accompanied
by momentum transfer and the atoms are allowed to move freely.
Since the transfer is coherent and reversible it serves as a 
basis for a hybrid
light-matter-wave Sagnac interferometer.

The rotational phase shift in an optical Sagnac interferometer
is given by
\be
\Delta \phi_{\rm light} = \frac{4\pi}{\lambda c} {\mathbf \Omega}\cdot 
{\mathbf A},
\label{Sagnac}
\ee
where ${\mathbf\Omega}$ and ${\mathbf A}$ are the vectors of the
angular velocity and the loop area. 
Discussing Fresnel dragging in EIT media 
Leonhardt and Piwnicki suggested in 
\cite{Leonhardt-PRA-2000} that the reduction 
of the group velocity in a solid attached to the
rotating body should lead to the replacement of the vacuum speed of light
$c$ in (\ref{Sagnac}) by the group velocity $v_{\rm gr}$ and thus to 
an enormous increase of rotational sensitivity.  
However, as will be shown here, the Sagnac phase shift in a
{\rm solid} medium is always given by the vacuum expression irrespective
of the group velocity. In fact in the history of 
Sagnac interferometry there had been a longer discussion about 
the effect of refractive materials in the beam path of passive optical
devices. It was finally recognized that materials that change the phase 
velocity of light have no effect on the Sagnac phase shift
(see e.g. \cite{Stedman-RPP-1997}).
However, if the coherence transfer from light to medium, 
associated with any group-velocity reduction, is accompanied by a 
{\it momentum transfer to freely moving particles} a traveling matter-wave
component is created which can lead to a substantial enhancement 
of the rotational sensitivity. 

\begin{figure}[ht]
\includegraphics[width=7.5cm]{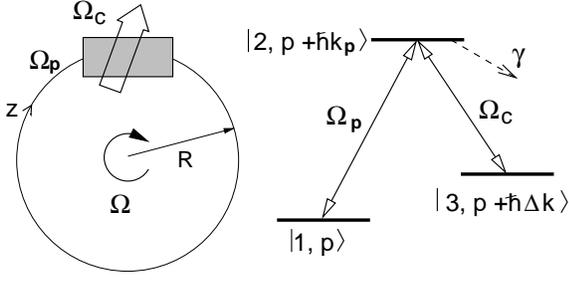} 
\caption{{\it left:} setup of light Sagnac interferometer with
vapor cell or trap attached to rotating body with angular velocity $
{\Omega}$. {\it right:} level scheme of atoms. $p$ denotes momentum
along the peripheral direction $z$. $k_p$ is the wavenumber of the probe
field propagating parallel to $z$. $\Delta k= k_p-k_c^\parallel$, where
$k_c^\parallel$ is the component of the control-field wave-vector in 
$z$ direction.}
\label{fig1}
\end{figure}

Let us consider a circular light interferometer of radius $R$
with an atomic vapor cell or trap in the beam path as shown in 
Fig.~\ref{fig1} attached to a rotating body.
The product of angular velocity $\Omega$ and radius $R$ 
is assumed to be sufficiently small
compared to $c$ such that non-relativistic quantum mechanics applies
\cite{Anandan-PRD-1981}.  
Light propagation as well as center-of-mass motion of the atoms shall 
be confined to the periphery of the loop, e.g. by means of optical fibers 
and appropriate atom potentials. Distances along the periphery will be 
denoted by the coordinate $z$.
As indicated in Fig.~\ref{fig1} the atoms shall have three 
internal states $|1\rangle,|2\rangle,|3\rangle$
allowing for a $\Lambda$-type
Raman coupling of the  $|1\rangle - |2\rangle$ transition to 
the probe field described by the Rabi-frequency $\Omega_p(z,t)$
and of the $|2\rangle - |3\rangle$ transition to 
the control field, characterized by the Rabi-frequency $\Omega_c$.
For the present application $\Omega_c$ is
assumed to be much larger than $\Omega_p$.


Under two-photon resonance conditions, the control field generates
electromagnetically induced transparency for the probe field 
associated with a reduction of the group
velocity according to 
$
v_{\rm gr}= c \, \cos^2\theta
$ 
with
$
\tan^2 \theta\equiv
{g^2 n}/{|\Omega_c|^2}\label{group1}
$
where $n$ is the density of atoms and $g=d 
\sqrt{\omega_p/(2\hbar\epsilon_0 F)}$ is a coupling constant 
containing the dipole moment $d$ 
of the probe transition and the transversal cross section $F$ of the
probe beam. As has been demonstrated in several
experiments \cite{slow-light} rather substantial 
group-velocity reductions can be achieved. 

The atoms are described by the matter fields $\Psi_1(z,t)$, $\Psi_2(z,t)$, 
and $\Psi_3(z,t)$ corresponding to the three internal states. 
Since the sources of the fields as well as the experimental apparatus
are attached to the rotating frame, the dynamics of the probe
light $\Omega_p$ 
and the matter-wave components $\Psi_\mu$
will be described in the rotating frame.
For this the total Hamiltonian in a non-rotating frame
$H$ will be transformed according to
$\exp\{i Gt\}H\exp\{-iGt\}$, where $G=\Omega {\hat M}$ 
is the generator of
a uniform rotation \cite{Bialynicki-Birula-PRL-1997}. 
$\hat M$ is the total (orbital) 
angular momentum of matter fields and light in the direction of 
${\mathbf\Omega}$. This leads to the following equations of motion
\be
{\cal D}\Psi_1 &=& 
\hbar \Omega_p^*\, {\rm e}^{i(\omega_p t-k_p z)}
\, \Psi_2,
\\
{\cal D}\Psi_2 &=& \hbar(\omega_2 -i\gamma)\Psi_2
+\hbar \Omega_p \,{\rm e}^{-i(\omega_p t-k_p z)}
\, \Psi_1
\\ 
&&
+\hbar\Omega_c {\rm e}^{-i(\omega_c t-k^\parallel_c z)} \Psi_3, \nn
\\
{\cal D}\Psi_3 &=& \hbar\omega_3 \Psi_3
+\hbar\Omega_c^*\, {\rm e}^{i(\omega_c t-k^\parallel_c z)} \Psi_2,
\ee
with
\be
{\cal D} &\equiv & i\hbar \partial_t+\frac{\hbar^2}{2 m}\partial_z^2 -i\hbar
\Omega R \partial_z.\nn
\ee
Here $\omega_p$ and $\omega_c$ are the probe and coupling frequencies in the
rotating frame, which is the laboratory frame,
and $k_p=\omega_p/c$ and $k_c^\parallel$ the 
corresponding wave-vector components in $z$ direction. 
$\gamma$ describes losses by spontaneous emission from the excited 
state $|2\rangle$. 
{\rm We assume that initially}, i.e. without applying 
the probe field all atoms are in state $|1\rangle$. 

We proceed by introducing slowly-varying
matter-wave amplitudes $\Psi_1= \Phi_1$, $\Psi_2=\Phi_2 {\rm e}^{-i(\omega_p t
-k_p z)}$, and $\Psi_3=\Phi_3
{\rm e}^{-i((\omega_p-\omega_c) t
-\Delta k z)}$, with $\Delta k=k_p-k_c^\parallel$ and
assume single-
and two-photon resonance of the carrier frequencies, i.e.
$\omega_p=\omega_2 + \hbar k_p^2/ 2m$ and $\omega_p-\omega_c
=\omega_3 +\hbar (\Delta k)^2/ 2m$.
Neglecting second-order derivatives within a slowly-varying
envelope approximation we arrive at the set of equations
\be
\Bigl(\partial_t-\Omega R\partial_z\Bigr)\Phi_1 &=& 
-i \Omega_p^* \Phi_2,\\
\Bigl(\partial_t-(\Omega R-v_{\rm rec})\partial_z\Bigr)\Phi_2 &=& 
+(i k_p \Omega R-\gamma)\Phi_2 \nn\\
&&-i\Omega_c \Phi_3 -i \Omega_p\Phi_1,\\
\Bigl(\partial_t-(\Omega R-\eta v_{\rm rec})\partial_z\Bigr)\Phi_3 &=& 
+i\eta k_p\Omega R\Phi_3\nn\\
&& -i\Omega_c^* \Phi_2,
\ee
where we have introduced the 
recoil velocity $v_{\rm rec}\equiv \hbar k_p/m$. The dimensionless
parameter $\eta\equiv \Delta k/k_p=1-k_c^\parallel/k_p$ describes 
the momentum transfer of
the light fields to the atoms. In a degenerate $\Lambda$ system
with co-propagating probe and control field there is no momentum transfer
and thus $\eta=0$.

Similarly we 
find the equation of motion of the probe field propagating 
parallel to the rotation in slowly-varying
envelope approximation
\be
\Bigl(\partial_t+ c\partial_z - i k_p\Omega R\Bigr)
\Omega_p(z,t)=-ig \Phi_2^*\Phi_1.
\label{Omegap}
\ee

We now make use of the assumption that the control field 
is strong compared to the probe field and treat the 
interaction with the fields in lowest order of perturbation
in $\Omega_p$. In this approximation the effect of the interaction on the
coupling field $\Omega_c$ can be disregarded. 
Furthermore we find for the matter field 
%
%
$\Phi_1(z,t) = \sqrt{n} = \textrm{const},
$
where $n$ is the density of atoms.
Assuming stationary 
conditions, the corresponding equations
for $\Phi_2$ and $\Phi_3$ read:
\be
&\Biggl[
\begin{array}{cc}
\Omega_c & -k_p \Omega R -i\gamma \\
-\eta k_p \Omega R & \Omega_c^*
\end{array}\Biggr]
\Biggl[
\begin{array}{c}
\Phi_3\\
 \Phi_2
\end{array}
\Biggr] = 
\Biggl[\begin{array}{c}
-g\sqrt{n}\, \Omega_p\\
0\end{array}
\Biggr]
&\nn\\
&+
\Biggl[\begin{array}{c}
i\bigl( v_{\rm rec} -\Omega R\bigr)
\partial_z  \Phi_2\\
i\bigl(\eta v_{\rm rec} -\Omega R\bigr)
\partial_z \Phi_3
\end{array}
\Biggr].
&\label{matrix}
\ee
If the complex amplitudes of the matter-wave components
change sufficiently slowly, we can analytically solve eqs.
(\ref{matrix}) by treating the term on the second line
as a perturbation. Substituting the corresponding result
into the equation for the probe field
(\ref{Omegap}) leads to the group velocity
\be
v_{\rm gr} = c \cos^2\theta + \eta v_{\rm rec} \sin^2\theta.\label{v-group}
\ee
One recognizes that the group velocity now contains an additional
term due to the momentum transfer associated with the coherence
transfer if $\eta\ne 0$ \cite{Dutton-Science-2001}. 
In the following we will assume that $\eta\ge 0$, although negative
values of $\eta$ are possible.  In the latter case the minimum
group velocity attainable is zero since the medium becomes
opaque as soon as $v_{\rm gr}$ crosses the value zero.
Thus if the mixing angle exceeds the critical value
%
%
$\tan^2\theta_{\rm crit} \equiv {c}/{v_{\rm rec}}
={m c^2}/{\hbar \omega_p}
$
%
%
the excitation propagates in the medium essentially as a matter-wave
with the recoil velocity. 

\begin{figure}[ht]
\includegraphics[width=6 cm]{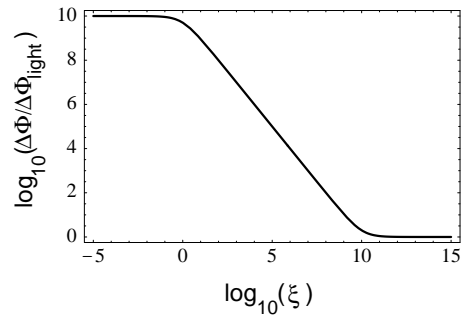} 
\caption{Sagnac phase-shift of the EIT hybrid interferometer relative to
the phase shift of an optical gyroscope of the same area and $\eta=1$; 
for $\xi\gg \eta mc^2/\hbar\omega_p$ we
are in the light and for $\xi\ll \eta$ we are in
the matter-wave regime.
}
\label{fig2}
\end{figure}

Under stationary conditions and keeping only terms up to first order
in the rotation velocity $\Omega$, one finds 
the simple probe-field equation
\be
\partial_z \ln \Omega_p(z) = 
i \frac{2\pi}{\lambda c} \Omega R
\frac{\xi(z)+ \eta\frac{m c^2}{\hbar\omega_p}}
{\xi(z)+\eta}, \label{E-simple}
\ee
where $\lambda$ is the probe-field wavelength
and we have introduced the parameter 
\be
\xi(z)\equiv \frac{\tan^2\theta_{\rm crit}}{\tan^2\theta(z)}
=\frac{v_{\rm gr}(z)}{v_{\rm rec}} -\eta,
\label{para-xi}
\ee
where in the last equation $v_{\rm gr}\ll c$ was assumed. (Note that
$v_{\rm gr}\ge \eta v_{\rm rec}$.)
Eq.(\ref{E-simple}) describes a simple phase shift without absorption
losses due to perfect EIT.
Two counter-propagating
probe fields will thus experience the Sagnac shift
%
%
\be
\Delta\phi & =&\frac{2\pi\Omega R}{\lambda c}\int{\rm d}z\, \frac{\xi(z)}{
\xi(z)+\eta}\nn\\
&&+\frac{\Omega R}{\hbar/m} \int{\rm d}z\, \frac{\eta}{\xi(z)+\eta}
.\label{phi}
\ee
The integration over $z$ takes into account that the group velocity
can be different in different parts of the interferometer loop
(see Fig.~\ref{fig1}).
If $\xi\gg \eta mc^2/\hbar\omega_p$, i.e. for $v_{\rm gr}\gg v_{\rm rec} 
mc^2/\hbar\omega_p$, 
eq.(\ref{phi}) reproduces the
Sagnac phase shift of an optical gyroscope, 
eq.(\ref{Sagnac}). On the other hand if $\eta\ne 0$ and $\xi\ll \eta$
(\ref{phi}) approaches the matter-field phase shift. 
Furthermore one recognizes that
in the absence of momentum transfer, i.e. for $\eta=0$, the phase shift is
for all values of $\xi$ identical to that of a light
interferometer. Thus the reduction of the group velocity alone does not
affect the Sagnac phase shift. One should note that this is in 
contrast to the strong Fresnel dragging always present in a medium
with a small group velocity. The difference to
Fresnel dragging arises because in the Sagnac interferometer there is no
motion of the medium relative to the source of the waves and consequently 
no simple Doppler-shift. In order to distinguish rotational phase shifts
from those resulting from linear accelerations it is necessary to 
use a symmetric interferometer setup. 
Since the matter wave is generated
by a coherent transfer from light  we can 
combine the rotational sensitivity per unit
area with a large interferometer area utilizing 
well established guiding techniques for
light. This is the main result of the present paper. The enhancement of the
Sagnac phase shift due to an atomic vapor with EIT put into the
beam path of a given optical interferometer is shown in Fig.~\ref{fig2}. 
For simplicity it is assumed that the vapor fills the whole volume of the
optical beam path.

In order to estimate the potential enhancement of rotational sensitivity
we now have to discuss the limitations on the minimum value of $\xi$, which
arise mainly from two-photon Doppler-shifts and velocity changing 
collisions.
As seen above, an enhancement of the rotational sensitivity requires
$\eta> 0$. In this case there is a two-photon Doppler-shift
caused by thermal motion of atoms in the rotating frame, which
leads to a finite absorption of the medium. As the absorption grows
with decreasing group velocity, this represents an essential limitation
to the hybrid interferometer. 
To discuss the effects of a thermal velocity distribution 
we decompose the three matter fields into velocity classes
according to
\be
\Phi_\mu(z,t) &=& \sum_v \Phi^v_\mu(z,t)\, {\rm e}^{i (q z - \nu t)}
\label{velocity-classes}
\ee
where $\hbar q = m v$ is the initial velocity of the atoms 
and $\hbar \nu= mv^2/2 - mv \Omega R$. We then can proceed as above
and obtain the same equations of motion for the slowly-varying 
velocity components
$\Phi_\mu^v$ as in eqs.(\ref{matrix}) except for the replacement
$\Omega R \longrightarrow \Omega R -v$. 

Keeping again only linear contributions of the angular velocity $\Omega$
to the phase, we find the following propagation equation for the probe
field
\be
&&\partial_z \ln \Omega_p(z) = i k_p \frac{\Omega R}{c} 
\frac{\xi(z)+\eta \frac{m c^2}{\hbar\omega_p}}{\xi(z) +\eta}
\nn\\
&&\qquad -2 ik_p\left(\frac{\gamma \eta k_p c}{g^2 n}
\frac{\overline{v^2}}
{v_{\rm rec}^2}\frac{1}{\xi(z)\bigl(\xi(z)+
\eta \bigr)}\right)^2\qquad\label{E-v}\\
&&\qquad - k_p \frac{\gamma  \eta k_p c}{g^2 n}
\frac{\overline{v^2}}{v_{\rm rec}^2} 
\frac{1}{\xi(z)\bigl(\xi(z)
+\eta\bigr)},\nn
\ee
which has the simple solution 
\be
\Omega_p(z) = \Omega_p(0) \, {\rm e}^{i\phi(z)}\, {\rm e}^{-\kappa z}.
\ee
In eq.~(\ref{E-v}) $\overline{v^2}$ denotes the mean square of the velocity 
distribution with $\overline{v}=0$. 

\begin{figure}[ht]
\includegraphics[width=8.5 cm]{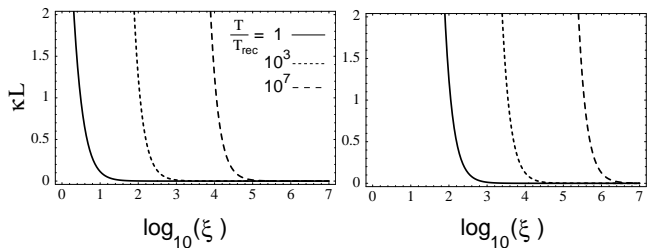} 
\caption{{\it left:} absorption coefficient for an 
atomic sample with typical trap size  ($L=
100\mu$m) and with opacity $\alpha=10^2$
{\it right:} absorption coefficient for an 
atomic sample with typical size of vapor cell $L=1$cm and $\alpha=10$.
$\lambda=500$nm}
\label{fig3}
\end{figure}

The first term in eq.~(\ref{E-v}) gives the Sagnac phase shift as
derived previously, the second term is a constant phase contribution
which however cancels in the difference phase of the two
counter-propagating waves and is thus
without consequence. The important new contribution due to the atomic 
velocity distribution is the third term in (\ref{E-v}) which gives
rise to absorption which grows with decreasing
$\xi$.
Thus $\xi$, resp. the group velocity, cannot be made arbitrarily small
since the growing losses
will lead to a decreasing signal-to-noise ratio. Whether or not 
the matter-wave limit $\xi\ll \eta$ can be reached will depend
on the velocity spread or temperature of the atoms. 
In deriving eq.~\ref{E-v} we have made furthermore the 
following simplifying assumptions 
$
\eta k_p^2 \overline{v^2}  \ll |\Omega_c|^2, 
\enspace{|\Omega_c|^4}/{\gamma^2}.
$
These are however well justified as long as the total amplitude 
decrease due to absorption is
less than e$^{-1}$ and
$\xi\ge\alpha^{-1}$, where $\alpha\equiv g^2 n z/\gamma c$ 
is the opacity of the medium 
in the absence of EIT, which is typically in the range between $1$ and 
$10^2$. Fig.~\ref{fig3} shows the absorption coefficient 
\be
\kappa L 
=\frac{(k_p L)^2}{\eta\alpha} 
\frac{T}{T_{\rm rec}}
\frac{1}{\xi(\xi+1)}.
\label{SNR}
\ee
for 
three different temperatures $T$ 
of the atomic
sample relative to the recoil temperature
$T_{\rm rec}$. 
In the left figure $\alpha=100, \lambda=500$nm, and $L=100\mu$m,
which corresponds to a typical situation in an ultra-cold atomic gas in
a trap. In the right figure $\alpha=10, \lambda=500$nm, and $L=1$cm,
corresponding to a typical vapor cell. 
One recognizes that by approaching the matter-wave regime the
absorption experiences a steep increase at a certain value of $\xi$.
Thus in order to retain a reasonable signal-to-noise ratio, i.e.
for absorption coefficients of $\kappa L\le 1$, there is a minimum
value of $\xi$ (resp. a minimum value of $v_{\rm gr}$).  
Fig.~\ref{fig3} shows that it is only possible to take advantage of the
maximum possible sensitivity enhancement of $10^{11}$
if the atomic sample is cooled down to the recoil limit or even
below. Nevertheless it is still possible to gain
a few orders of magnitude compare to the pure
light regime by moderate cooling ($T \approx 10^3\, T_{\rm rec}$).

Another important limitation to the minimum value of $\xi$
is set by the decoherence of motional degrees of freedom
of the matter-wave component caused by collisions.
Although superpositions of internal states can survive
very many collisions, in particular if they
are hyperfine components of the electronic ground state, velocity
kicks quickly destroy the coherence between motional
states. To  account for the latter
in a quantitative way a kinetic theory needs to be developed.
Such a theory goes however beyond the scope of the present paper 
and will be the subject of future work. We here can only give 
an upper bound for the corresponding limit on $\xi$ (or equivalently 
the group velocity $v_{\rm gr}$) by requiring that the
pulse delay time $\tau_{\rm d}$ is shorter than the
average time $\tau_{\rm coll}$ between two successive 
velocity-changing collisions:
\be
\left(\frac{v_{\rm gr}}{v_{\rm rec}}\right)_{\rm min}
\le \frac{L}{v_{\rm rec}\tau_{\rm coll}}=L n \sigma 
\sqrt{\frac{T}{T_{\rm rec}}}\label{coll-limit}
\ee
where $\sigma$ is the collisional cross section. E.g. in a gas cell 
of $L=1$cm, $n=10^{11}$cm$^{-3}$, assuming 
$\sigma=10^{-12}..10^{-10}$cm$^2$ yields $v_{\rm gr}^{\rm min}/v_{\rm rec}
\approx 0.1 ... 10 \sqrt{T/T_{\rm rec}}$.
When a {\it coherent}
sample of atoms such as a BEC is used, the limitation due to collisions
is of course absent.

In the present paper we have shown that it is possible to combine
the advantages of light  and mater-wave gyroscopes, to measure
the Sagnac phase-shift making use of the coherence and momentum
transfer associated with ultra-slow light propagation in
cooled atomic vapors. 

\

M.F. would like to acknowledge discussions with 
L.J. Wang, R. Chiao and U. Leonhardt during the KITP workshop on 
``Fast and slow light and meta-materials'' at the UC Santa Barbara
as well as with I. Bialynicki-Birula.



\end{document}